\documentclass[12pt,a4paper]{iopart}
\usepackage{siunitx, color, graphicx,subcaption,url,booktabs}
\usepackage{tablefootnote}

\usepackage{csquotes}
\usepackage{textcomp,lineno,booktabs,amssymb} 
\begin{document}

\title[]{Feasibility study of a proton CT system based on
4D-tracking and residual energy
determination via time-of-flight}
\author{Felix Ulrich-Pur$^1$, Thomas Bergauer$^1$, Alexander Burker$^2$, Albert Hirtl$^2$, Christian Irmler$^1$, Stefanie Kaser$^1$, Florian Pitters$^1$, Simon Rit$^3$}
        \address{$^1$ Austrian Academy of Sciences, Institute of High Energy Physics (HEPHY),
        Nikolsdorfer Gasse 18, 1050 Wien, Austria}
		\address{$^2$ TU Wien, Atominstitut, Stadionallee 2, 1020 Wien, Austria}
		\address{$^3$ Lyon University, INSA-Lyon, University Lyon1, UJM-Saint Etienne, CNRS, Inserm, CREATIS UMR5220, U1206, France}
		\ead{felix.ulrich-pur@oeaw.ac.at}
\vspace{10pt}
\begin{indented}
\item[]September 2021
\end{indented}
\begin{abstract}
For dose calculations in ion beam therapy, it is vital to accurately determine the relative stopping power (RSP) distribution within the treated volume.  Currently, RSP values are extrapolated from Hounsfield units (HU), measured with x-ray computed tomography (CT), which entails RSP inaccuracies due to conversion errors. A suitable method to improve the treatment plan accuracy is proton computed tomography (pCT). A typical pCT system consists of a tracking system and a separate residual energy (or range) detector to measure the RSP distribution directly. This paper introduces a novel pCT system based on a single detector technology, namely low gain avalanche detectors (LGADs). LGADs are fast 4D-tracking detectors, which can be used to simultaneously measure the particle position and time with precise timing and spatial resolution. In contrast to standard pCT systems, the residual energy is determined via a time-of-flight (TOF) measurement between different 4D-tracking stations.\\ The design parameters for a realistic proton computed tomography system based on 4D-tracking detectors were studied and optimized using Monte Carlo simulations. The RSP accuracy and RSP resolution were measured inside the inserts of the CTP404 phantom to estimate the performance of the pCT system. After introducing a dedicated calibration procedure for the TOF calorimeter, RSP accuracies $<\SI{0.6}{\percent}$ could be achieved. Furthermore, the design parameters with the strongest impact on the RSP resolution were identified and a strategy to improve RSP resolution is proposed.
\end{abstract}
\noindent{\it Time-of-flight, 4D-tracking detector, Low Gain avalanche detectors, proton computed tomography, proton therapy, RSP accuracy, RSP precision\/}\\
\submitto{\PMB}

 \section{Introduction}
Ion beam therapy is used to treat deep-seated tumours whilst minimizing normal-tissue damage as much as possible \cite{Linz2012} due to the highly localized depth-dose characteristics of ions in matter. As the ion traverses the patient, the energy deposition per unit path length, referred to as stopping power (SP), increases until the particle is at rest, which leads to a maximum of deposited energy at the end of the particle's range (Bragg peak). Since the location of the Bragg peak strongly depends on the SP of the traversed material \cite{braggkleemann}, it is crucial to accurately determine the SP distribution within the patient to reduce any range uncertainties, which would require larger safety margins around the treated volume \cite{Yang_2012}. Currently, treatment plans are based on single energy x-ray computed tomography (SECT) scans, where the obtained Hounsfield units (HU) are converted to relative stopping power (RSP), i.e. SP expressed relative to water \cite{Schaffner_1998}. This extrapolation from HU to RSP introduces additional uncertainties, which, depending on the tissue, leads to RSP errors ranging from $\SI{1.6}{}$ to $\SI{5}{\percent}$ \cite{Yang_2012}, or for more modern dual-energy CT (DECT) scanners $\gtrapprox\SI{0.6}{\percent}$ \cite{Wohlfahrt2017,Hudobivnik2016,Dedes2019}.
\\ Alternatively, to improve the RSP accuracy even further ($\approx \SI{0.1}{\percent}$), proton computed tomography (pCT) could be used to measure the RSP distribution directly \cite{Poludniowski_2014}. In pCT, the particle's path through the patient and the deposited energy inside the patient are measured for each particle and are used to reconstruct the 3D RSP map \cite{schulte2004}. A typical pCT scanner consists of two tracker pairs placed upstream and downstream of the patient to measure the particle's position and direction. To measure the energy loss inside the patient, either a residual range or residual energy detector is placed downstream of the rear tracker pair. Different pCT systems \cite{Scaringella_2014,Pettersen2019,Esposito2018,SADROZINSKI2013205} have been developed in recent years, showing the potential to compete with modern DECT scanners \cite{Dedes2019}. As described in \cite{schulte2004}, a pCT system suitable for clinical use should be able to measure the RSP with an accuracy $\leq \SI{1}{\percent}$ and a spatial resolution $\leq \SI{1}{mm}$. In order to keep the acquisition time of a full pCT scan comparable to a normal CT scan ($<\SI{1}{min}$), data acquisition rates of at least a few $\SI{}{MHz}$ are required. Fulfilling all these requirements while keeping the production and maintenance costs as well as the system's complexity as low as possible proves to be challenging. \\A possible solution for a clinically applicable pCT system could be based on 4D-tracking detectors used for both particle path estimation and time-of-flight (TOF) residual energy measurements \cite{Vignati_2020}. Low gain avalanche detectors (LGADs), for example, are promising candidates since they have high rate-capabilities and offer timing resolutions in the order of $\SI{30}{}-\SI{50}{ps}$ \cite{CARTIGLIA201783,Pietraszko2020} and, depending on the LGAD technology, can have spatial resolutions down to few tens of $\SI{}{\micro m}$ \cite{Tornago2021,Currs2020}. \\
The aim of this work is to present a comprehensive feasibility study of an LGAD-based pCT system, which should serve as a guide for future hardware developments. For that purpose, the influence of various detector design aspects on the performance of the TOF-pCT scanner has been studied and is presented in two parts. First, the impact of different system parameters of a stand-alone TOF calorimeter on the energy resolution and accuracy of the residual energy measurement is explored. Also, a dedicated calibration procedure for the TOF calorimeter is presented. Second, the performance of an LGAD-based TOF-pCT system using the same detector technology for particle tracking and residual energy determination is investigated. The RSP accuracy and precision are measured using the CTP404 phantom and are then compared to the results of the latest pCT scanner \cite{Dedes2019}.\\

\section{Materials and methods}
To assess the performance of the investigated pCT systems based on 4D-tracking detectors, Monte Carlo (MC) simulations of realistic TOF-pCT systems were performed and compared to simulations of an ideal pCT setup, without a TOF calorimeter and with ideal energy and position measurement. All pCT setups were modeled in Geant4 (version 10.05.1) \cite{AGOSTINELLI2003250} using the \textit{QGSP\_BIC\_EMY} physics list with \textit{EM\_Options 3}. 
\subsection{Time-of-flight calorimeter}  \label{sec:tofcalmeth}
First, basic design choices for the TOF-pCT system were made based on separate MC simulations of a realistic stand-alone TOF calorimeter. To estimate the performance of each of the investigated calorimeter settings,  the energy resolution and absolute error were determined as a measure for precision and accuracy.
\subsubsection{Setup geometry}
\begin{figure}[!h]
    \begin{center}
    \includegraphics[width=0.95
   \textwidth]{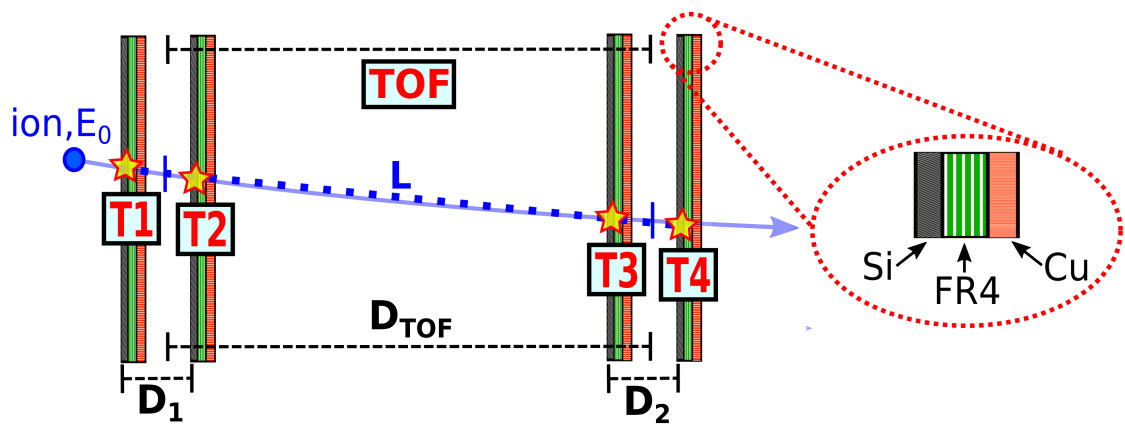}
    \end{center}
    \caption{Schematic of a TOF calorimeter based on LGAD detectors. The TOF is measured between two timing stations (T1/T2 and T3/T4), each consisting of two generic LGAD planes, modelled as a silicon (Si), copper (Cu) and flame retardant glass epoxy (FR4) compound (the cross section of an LGAD plane is depicted on the right). A straight line fit through all hit positions $(x_i,y_i)$ is used to estimate the flight path length $L$ (dotted line).}
    \label{fig:lgadsketch}
    \end{figure} 
\mbox{}\\A schematic drawing of the simulated setup is depicted in figure \ref{fig:lgadsketch}. The TOF calorimeter was simulated using two timing stations, each consisting of $n$=2 4D-tracking planes, placed $\SI{10}{cm}$ apart ($\mathrm{D}_1$=$\mathrm{D}_2$=$\SI{10}{cm}$). To create a realistic model of a TOF calorimeter, each tracking plane was modelled as a generic LGAD module, consisting of a silicon layer (Si) for the sensor and a copper (Cu) and flame retardant glass epoxy (FR4) layer to model a printed circuit board (PCB). The thickness $\mathrm{X}$ of each material inside an LGAD plane ($\mathrm{X}^{\mathrm{Si}}$, $\mathrm{X}^{\mathrm{Cu}}$, $\mathrm{X}^{\mathrm{FR4}}$) was then varied to simulate different detector technologies (e.g. strip sensors or pixel detectors). To compare the material composition of different detector technologies, the material budget \cite{Gupta:1279627} per tracking plane was calculated using 
  	\begin{equation}
  	\epsilon= \sum_i \frac{\mathrm{X}^i}{{\mathrm{X}_0^i}}=\frac{\mathrm{X}^{\mathrm{Si}}}{{\mathrm{X}_0^\mathrm{Si}}}+\frac{\mathrm{X^\mathrm{Cu}}}{{\mathrm{X}_0^\mathrm{Cu}}}+\frac{\mathrm{X}^{\mathrm{FR4}}}{{\mathrm{X}_0^\mathrm{FR4}}}.
  	\end{equation}
The radiation lengths for Si (${\mathrm{X}_0^\mathrm{Si}}$=$\SI{93.7}{mm}$) and copper (${\mathrm{X}_0^\mathrm{Cu}}$=$\SI{14.36}{mm}$) were extracted from \cite{PDG} and FR4 ($\mathrm{X}_0^\mathrm{FR4}$=$\SI{167.608}{mm}$) from \cite{FR4}. Standard LGAD sensors typically consist of a $\SI{300}{}$-$\SI{350}{\micro m}$ thick silicon layer \cite{CARULLA2019373}, which corresponds to a material budget of $\approx\SI{0.32}{}$-$\SI{0.37}{\percent}$ $\mathrm{X}/\mathrm{X}_0$. Depending on the detector technology, the silicon sensor can also be mounted on a PCB. For example, the LGAD pad detectors, as described in \cite{CARTIGLIA201783} were mounted on the USCS readout board \cite{uscsboard} and had a total material budget of $\approx\SI{2}{\percent}$ $\mathrm{X}/\mathrm{X}_0$. Within our studies, three different material budgets were simulated ($\SI{0.1}{\percent}$, $\SI{1}{\percent}$ and $\SI{2.3}{\percent}$ $\mathrm{X}/\mathrm{X}_0$) to cover a larger range of possible values. $\SI{0.1}{\percent}$ $\mathrm{X}/\mathrm{X}_0$ and $\SI{2.3}{\percent}$ $\mathrm{X}/\mathrm{X}_0$ are more extreme examples and correspond to a single $\SI{100}{\micro m}$ silicon layer and a compound consisting of a $\SI{300}{\micro m}$ silicon, $\SI{200}{\micro m}$ copper and $\SI{1}{mm}$ FR4 layer.\\
To model the intrinsic spatial and timing resolution of a realistic LGAD sensor, the transversal hit position ($x_i,y_i$) and the hit time $\mathrm{T}_i$ at sensor $i$ were blurred using a Gaussian distribution with a standard deviation of $\sigma_{xy}$ and $\sigma_{\mathrm{T}}$, respectively. For all settings, the spatial resolution $\sigma_{xy}$ was varied between $\SI{0}{}$ and $\SI{5}{cm}$ to study the influence of the spatial granularity of different sensor geometries. Also different intrinsic time resolutions $\sigma_{\mathrm{T}}$, ranging from $\SI{0}{}$ to $\SI{100}{ps}$ were investigated.
\subsubsection{Residual energy determination via time-of-flight measurements}\label{sec:emeas}\mbox{}\\
For the TOF-based energy measurement, the mean hit time per timing station was used to calculate the TOF through the calorimeter according to 

\begin{equation}
\mathrm{TOF}= \frac{1}{n}\left(\sum_i^n \mathrm{T}_{i+n}- \mathrm{T}_i \right),
\label{eq:toa}
\end{equation} 
with $n=2$ LGADs/timing station. Assuming a constant velocity $v$ along the particle path inside the TOF calorimeter, the kinetic energy of each particle with mass $m_0$ was calculated as follows
\begin{equation}
E_{\mathrm{kin}}=m_0 c^2\cdot\left(\frac{1}{\sqrt{1-\frac{{v}^2}{c^2}}}-1\right), 
\label{eq:evstof}
\end{equation} 	
using $v\approx L/\mathrm{TOF}$, where $L$ is the flight path length (figure \ref{fig:lgadsketch}) and TOF is the time-of-flight as defined in equation (\ref{eq:toa}). For the path length estimation, a straight-line fit through all hit positions ($x_i,y_i$) along the particle's path was used (dotted line in figure \ref{fig:lgadsketch}). The distance between the points on the resulting straight line, located at the centre between each timing station, was used to calculate the flight path length $L$.\\
To study the influence of the TOF calorimeter length, the flight distance $\mathrm{D}_{\mathrm{TOF}}$ was varied between $\SI{0.5}{}$ and $\SI{2}{m}$. The performance of each calorimeter setting was tested using $\SI{1e5}{}$ primary protons with kinetic energies ranging from $\SI{50}{}$ to $\SI{400}{MeV}$. 
\subsubsection{Precision of the residual energy measurement}\label{sec:eprec}
  	\mbox{}\\
  	To estimate the precision of the residual energy measurement, the energy resolution $\sigma_{E_{\mathrm{meas}}}/E_{\mathrm{meas}}$, was determined via simulation for each setting. In addition, the theoretical energy resolution was calculated using first order Gaussian error propagation (GEP) of equation (\ref{eq:evstof})
  	\begin{equation}
\frac{\sigma_{E_{\mathrm{kin}}}}{E_{\mathrm{kin}}} = \frac{1}{E_{\mathrm{kin}}} \cdot\sqrt{{\left(\frac{\partial E_{\mathrm{kin}}}{\partial \mathrm{TOF}} \cdot \sigma_{\mathrm{TOF}} \right)}^2+{\left(\frac{\partial E_{\mathrm{kin}}}{\partial L}\cdot \sigma_{L} \right)}^2}.
\label{eq:toferr}
\end{equation} 
The resulting theoretical energy resolution
\begin{equation}
\frac{\sigma_{E_{\mathrm{kin}}}}{E_{\mathrm{kin}}} = \frac{ {\gamma}^3 {\beta}^2}{L \left(\gamma -1 \right)} \sqrt{{{\sigma}_{\mathrm{TOF}}}^2 {\beta}^2 c^2+{\sigma_{L}}^2},
\label{eq:eres}
\end{equation}
with $\gamma=1/\sqrt{1-\beta^2}$ and $\beta=v/c$, is equivalent to the energy resolution described in \cite{volzphd2021}, if no path length uncertainty ($\sigma_L=0)$ is assumed. To calculate the uncertainty of the total TOF (${\sigma}_{\mathrm{TOF}}$), a GEP of equation (\ref{eq:toa}) was performed, resulting in
\begin{equation}
   	{\sigma}_{\mathrm{TOF}}= \sqrt{\sum_i^{2n} {\left(\frac{\partial \mathrm{TOF}}{\partial {\mathrm{T}}_i} \cdot {\sigma}_{{\mathrm{T}}_i} \right)}^2}=\sqrt{\frac{2}{n}} \cdot {\sigma}_{\mathrm{T}}.
   	\label{eq:planes}
   	\end{equation}
Since the same intrinsic time resolution per LGAD was assumed ($\sigma_{\mathrm{T}_i}={\sigma}_{\mathrm{T}}$) and two LGADs were used per timing station ($n=2$), the theoretical uncertainty of the TOF is equivalent to the intrinsic time resolution per plane. For each setting, the energy resolution was obtained from a simulation of a realistic setup and compared to the theoretical energy resolution (equation (\ref{eq:eres})).
\subsubsection{Accuracy and calibration of a realistic TOF calorimeter}
  	\mbox{}\\
  	In section \ref{sec:emeas} and section \ref{sec:eprec}, the energy loss of ions along their path through the TOF calorimeter was neglected. However, since the investigated setup is placed in air and consists of multiple LGAD detectors mounted on a PCB, a significant energy loss is expected. Therefore, to improve the accuracy of the energy measurement, it is important to account for the energy loss along the particle's trajectory and to apply a suitable energy calibration. For each setting, the primary beam energy $E_{\mathrm{in}}$ was varied between $\SI{50}{}$ and $\SI{400}{MeV}$ and the absolute error of the energy measurement $\Delta E$, was calculated according to
  	\begin{equation}
  	\Delta E = E_{\mathrm{meas}}-E_{\mathrm{in}} .
  	\label{eq:abserr}
  	\end{equation}
  Then, a function $f\left( E_{\mathrm{meas}} \right)$, as described in section \ref{sec:resultcalib}, was fitted to equation (\ref{eq:abserr}) to approximate the absolute error $\Delta E\approx f\left( E_{\mathrm{meas}} \right)$. The function $f\left( E_{\mathrm{meas}} \right)$ is an empirical model taking into account the energy loss inside the calorimeter as well as inaccuracy of the energy measurement due to the non-linear relation between energy and TOF (see equation (\ref{eq:evstof}) and section \ref{sec:resultcalib}). The resulting calibration curves were used for all following simulations to correct the inaccuracies of the energy measurement as follows
  	\begin{equation}
  	E_{\mathrm{in}}\left( E_{\mathrm{meas}} \right) \approx E_{\mathrm{meas}}+f\left( E_{\mathrm{meas}} \right).
  	\label{eq:calibcurve}
  	\end{equation}
After applying the calibration according to equation (\ref{eq:calibcurve}), the relative error
\begin{equation}
    \epsilon_E=|E_{\mathrm{in}}-E_{\mathrm{meas}}|/E_{\mathrm{in}}
    \label{eq:relerrenergy}
\end{equation} was calculated for each setting to estimate the accuracy of the energy measurement.

   \subsection{Proton computed tomography with 4D tracking detectors}
   \subsubsection{Experimental setup}
\begin{figure}[!h]
    \begin{center}
    \includegraphics[width=0.9\textwidth]{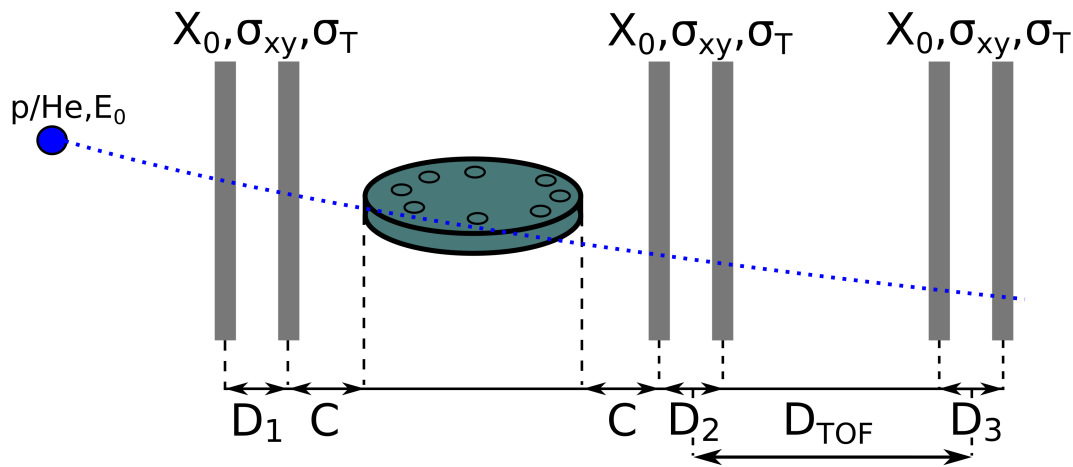}
    \end{center}
    \caption{Schematic overview of the simulated pCT setup based on 4D tracking detectors. The performance of the proposed pCT system was studied for different system parameters using the CTP404 phantom.}
    \label{fig:ictsetup}
    \end{figure}

\mbox{}\\
	After simulating a stand-alone TOF calorimeter (see section \ref{sec:tofcalmeth}), a full pCT system based on 4D tracking detectors was modelled in Geant4 (figure \ref{fig:ictsetup}). Different system parameters, as described in table \ref{tab:params}, were varied and optimized to design a pCT system that could potentially fulfil the clinical requirements as defined in \cite{schulte2004}. For each setup geometry, a calibration of the TOF calorimeter was performed prior to the pCT measurement (equation (\ref{eq:calibcurve})). The energy loss inside the first two tracking planes was calculated to correct the primary beam energy $E_0$, which is needed for the reconstruction.
\begin{table}[h]   
\begin{center}

\begin{tabular}{l @{\hskip 0.6cm}l @{\hskip 1.3cm}r   }
\toprule
 \textbf{parameter}&\textbf{name} & \textbf{simulated values}   \\
 \midrule
 $\mathbf{X}/\mathbf{X_0}$ &material budget & $\SI{0.1}{}$, $\SI{1}{}$, $\SI{2.3}{\percent}$ \\
 \boldmath{$\mathbf{\sigma_{xy}}$}  &spatial resolution &$\SI{0}{\micro m}$  \\
  \boldmath{$\mathbf{\sigma_{T}}$} &time resolution per tracking plane &$\SI{0}{}$, $\SI{10}{}$, $\SI{30}{}$, $\SI{50}{}$, $\SI{100}{ps}$ \\
  \textbf{C} &phantom clearance & $\SI{10}{cm}$  \\
  $\mathbf{D_{1,2,3}}$ &distance between tracking planes  & $\SI{10}{cm}$  \\
  $\mathbf{D_{TOF}}$ &flight distance & $\SI{50}{},\SI{100}{},\SI{150}{},\SI{200}{cm}$ \\
  $\mathbf{E_{0}}$ &beam energy  & $\SI{200}{}$, $\SI{250}{}$, $\SI{300}{}$, $\SI{350}{}$, $\SI{400}{MeV}$ \\
  \bottomrule

  \end{tabular}
  \end{center}

  \caption{Summary of the varied pCT system parameters to study the overall performance.}
  \label{tab:params}
\end{table}

\subsubsection{Phantom}
\begin{figure}[h]
 	\begin{center}
 	\includegraphics[width=0.4\textwidth]{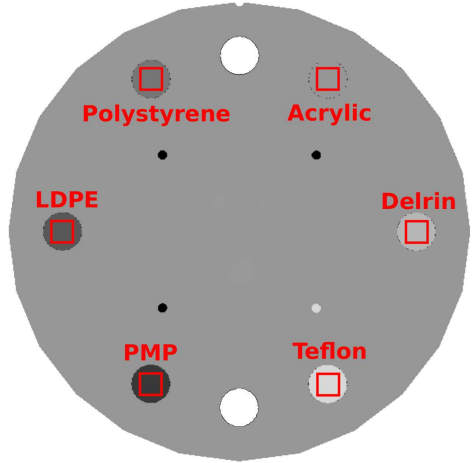}
 	\end{center}
 	\caption{The CTP404 phantom was used to study the performance of the proposed pCT system. The RSP was measured at six different inserts using square-shaped ROIs, highlighted as red squares in this figure.}
 	\label{fig:centralslice}
 	\end{figure}

\mbox{}\\\label{sec:phantom}
In order to compare the results of the proposed TOF-pCT system to the latest pCT scanners, the same phantom as in \cite{Dedes2019} was used to measure the performance in terms of RSP accuracy and precision. As depicted in figures \ref{fig:ictsetup} and \ref{fig:centralslice}, the CTP404 phantom is a cylindrical PMMA phantom with a diameter of $\SI{15}{cm}$ and a thickness of $\SI{2.5}{cm}$. Inside the PMMA cylinder, inserts with different materials, shapes and sizes are placed. Following the example of \cite{Dedes2019}, only the cylindrical inserts with a diameter of $\SI{12.5}{mm}$ were analysed. The investigated inserts consist of polymethylpentene (PMP), Teflon, polyoxymethylene (POM, also known as Delrin), Polystyrene, polyethylene (LDPE) and Acrylic. After reconstruction, the RSP was obtained in square-shaped regions of interest (ROIs) with a side length of $\SI{6}{mm}$ placed at the centre of each insert (figure \ref{fig:centralslice}). 
\subsection{Reconstruction and analysis}
\subsubsection{Forward projections and reconstruction}\mbox{}\\
For each of the investigated setup geometries, 360 projections were recorded at $\SI{1}{^\circ}$ steps and used to reconstruct the RSP map inside the phantom. Per projection, the phantom was irradiated with $\approx$$ \SI{100}{}$ primary particles per $\SI{}{mm^2}$. The obtained forward projections were further processed by applying standard 3-$\sigma$ cuts on kink angle and energy loss to eliminate large-angle scattering events \cite{Schulte2008}. Following the example of \cite{Dedes2019}, the pCT image was reconstructed using distance-driven-binning (DDB) \cite{Rit2013} and $\SI{1}{mm^3}$ voxels with a slice thickness of $\SI{1}{mm}$. DDB is a reconstruction algorithm for pCT which is based on filtered back-projection and approximates the particle's trajectory via most likely path (MLP) estimation \cite{Schulte2008}.

\subsubsection{RSP accuracy and precision}\label{sec:rsp}\mbox{}\\
Fifteen slices per insert were used to obtain the RSP distribution inside each insert. A Gaussian distribution was fitted to the RSP distribution to obtain the mean $\mu_{\mathrm{RSP}}$ and the standard deviation $\sigma_{\mathrm{RSP}}$ of the RSP. Since the RSP is very sensitive to the material definition in the simulation, reference RSP values $\mathrm{RSP}_\mathrm{ref}$ were obtained via separate MC simulations based on residual range measurements in water \cite{Michalak2017} using the same material composition and physics list as for the pCT simulation. For the residual range measurement, an absorber with thickness $t_m$ was placed in front of a water block, which was irradiated with protons with an energy $E$. This resulted in a shift in the residual range according to the water equivalent thickness (WET) \cite{Zhang2009} of the absorber
\begin{equation}
WET=\int \mathrm{RSP} \cdot dx \approx \mathrm{RSP} \cdot t_m.
\label{eq:wet}
\end{equation}
Then, the $\mathrm{RSP}_\mathrm{ref}$ was obtained for each material by dividing the measured WET by the thickness $t_m$ of the absorber. In order to estimate the shift in terms of range, the R20 (position in the water block, where the energy deposition has decreased to $\SI{20}{\percent}$ of its maximum) was used. 
\\
To study the performance of each of the investigated TOF-pCT setups, the coefficient of variation 
\begin{equation}
\label{eq:cvrsp}
\mathrm{CV}=\frac{\sigma_{\mathrm{RSP}}}{\mu_{\mathrm{RSP}}}
\end{equation} was used as an estimator for the RSP precision and the relative error 
\begin{equation}
 \label{eq:relerrrsp}
\epsilon_{\mathrm{RSP}}=\frac{|\mathrm{RSP}_\mathrm{ref}-\mathrm{RSP}_\mathrm{meas}|}{\mathrm{RSP}_\mathrm{ref}}
\end{equation} was calculated for each insert. To estimate the overall RSP accuracy of the pCT scan, the mean absolute percentage error (MAPE) was calculated according to
\begin{equation}
\mathrm{MAPE}=\frac{\sum_i^{n_{\mathrm{mat}}}\epsilon_{\mathrm{RSP,i}}}{n_{\mathrm{mat}}}, 
\end{equation}
with $n_{\mathrm{mat}}=6$ different inserts.

The measured CV and MAPE were then compared to an ideal pCT simulation, assuming infinitesimally thin sensors ($\mathrm{X}/\mathrm{X}_0=0$) and a perfect energy and position measurement. For the ideal setup, the input energy and the residual kinetic energy, which are both required for the reconstruction \cite{Rit2013}, were measured directly at the second tracker upstream of the phantom and the first tracker downstream of the phantom, respectively.\\

\section{Results}
\subsection{Time-of-flight calorimeter}
In this section, results showing the performance of a stand-alone TOF calorimeter for different system settings are presented. The goal is to identify and optimize the system parameters which dominate the energy measurement.
\subsubsection{Energy resolution}
\begin{figure}[h]
\begin{center}
\includegraphics[width=0.99\textwidth]{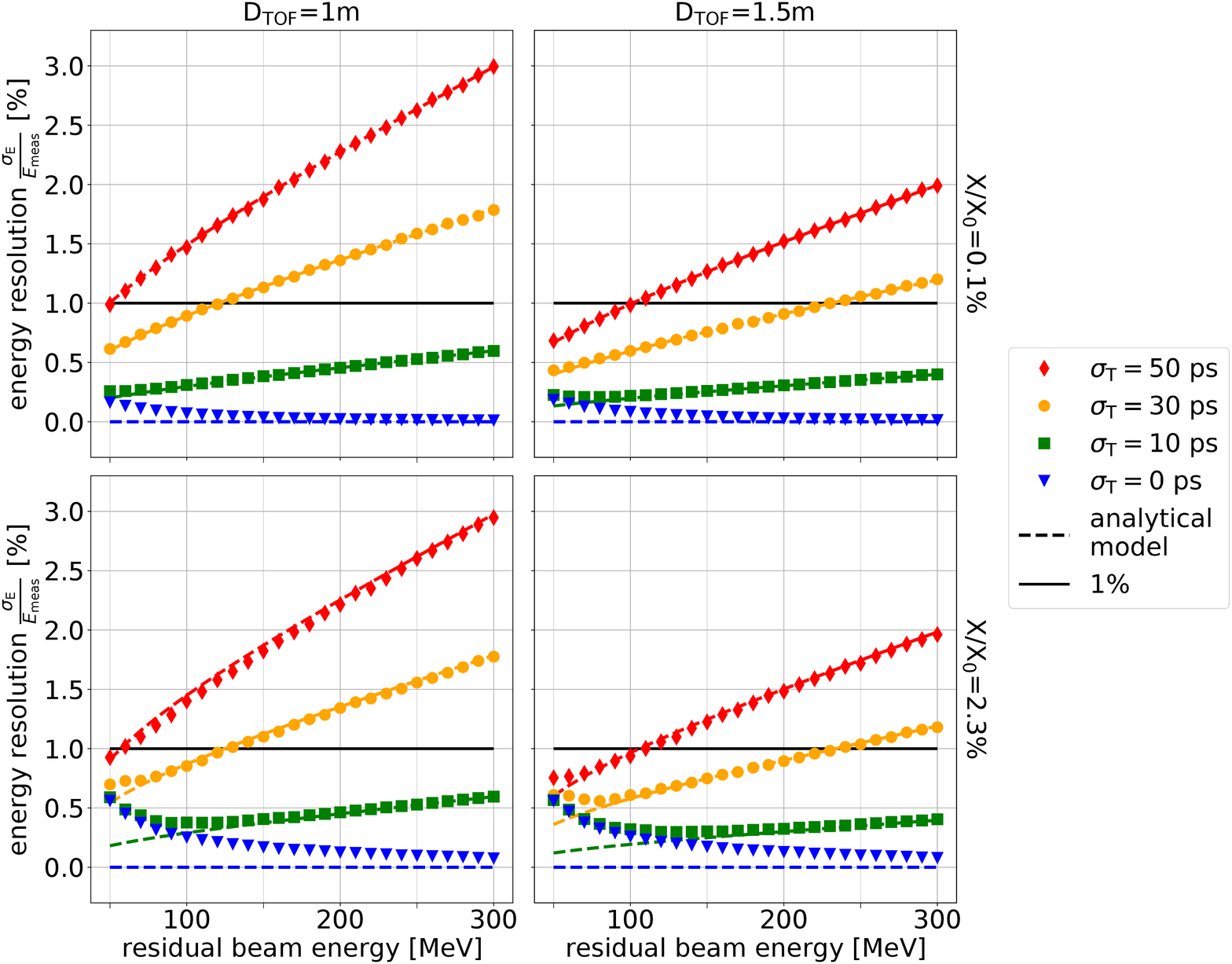}
\end{center}
\caption{The energy resolution was simulated (diamond-shaped, circular and square markers) for different system parameters and compared to the analytical model (dashed lines) assuming $\sigma_L=0$. Two material budgets, $\SI{0.1}{\percent}$ (top) and $\SI{2.3}{\percent}$ (bottom), and two flight distances, $\SI{1}{m}$ (left) and $\SI{1.5}{m}$ (right), were considered. For comparison, the $\SI{1}{\percent}$ requirement for a single-staged residual energy calorimeter \cite{Bashkirov2016} is also shown.}
\label{fig:eresresults}
\end{figure}

\mbox{}\\
Figure \ref{fig:eresresults} shows the energy resolution obtained from a simulation of a realistic TOF calorimeter compared to the analytical model as described in equation (\ref{eq:eres}). In addition, a simulation of an ideal TOF measurement with $\sigma_{\mathrm{T}}$= $\SI{0}{ps}$ was performed for each setting to estimate the impact of energy straggling inside the calorimeter. As can be seen in figure \ref{fig:eresresults}, energy resolutions close to the straggling limit could be observed for $\sigma_{\mathrm{T}}$= $\SI{10}{ps}$ and for beam energies $\leq\SI{100}{MeV}$. Also, for settings with more realistic intrinsic time resolutions $\geq \SI{30}{ps}$ and $\mathrm{X}/\mathrm{X}_0$=$\SI{2.3}{\percent}$, energy straggling along the particle path has to be taken into account if residual beam energies $\leq \SI{70}{MeV}$ are expected. For all other investigated settings, the precision of the energy measurement is well described by the analytical model as defined in equation (\ref{eq:eres}). According to equation (\ref{eq:eres}), the most dominating factors influencing the energy resolution are the intrinsic time resolution per tracking plane, flight distance and beam energy, which can also be observed in figure \ref{fig:eresresults}.

\subsubsection{Accuracy and calibration of a realistic TOF calorimeter}
\begin{figure}[!h]
  \begin{center}
  \includegraphics[width=0.99\textwidth]{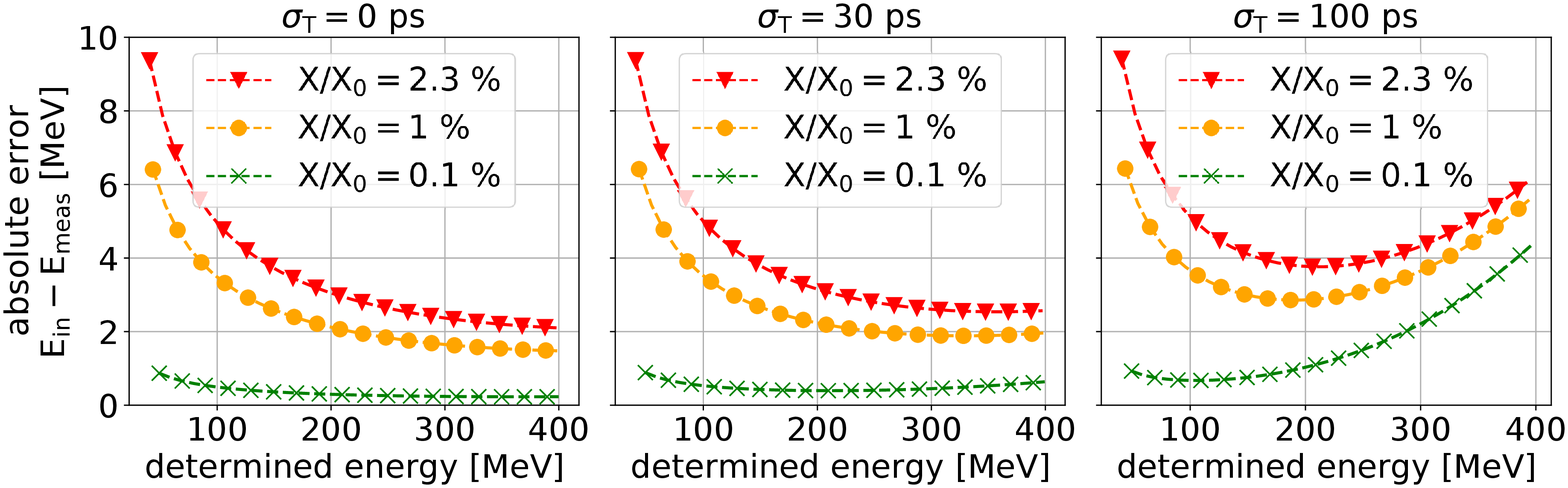}
 
\caption{Simulated absolute error of the energy measurement for an LGAD based TOF calorimeter with a flight distance of $\SI{0.5}{m}$, different material budgets, intrinsic time resolutions per tracker plane and determined residual proton energies ($E_{\mathrm{meas}}$). For each setting a calibration curve was obtained, shown as the dashed lines. }
\label{fig:praecalib}
\end{center}

\end{figure}

\begin{figure}[!h]
  \begin{center}
\includegraphics[width=0.8\textwidth]{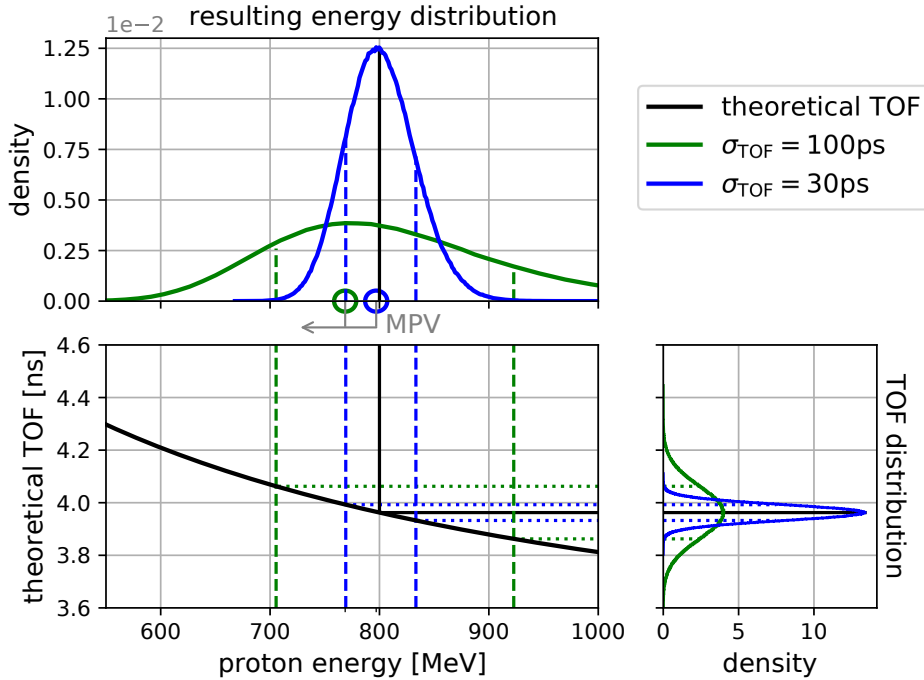}

\caption{The influence of the intrinsic timing resolution per tracking plane on the energy measurement is shown for $\SI{800}{MeV}$ protons and a TOF path length of $\SI{1}{m}$. The non-linear relation between TOF and energy (bottom left) results in an asymmetric shift of the energy distribution (top left) if high TOF uncertainties are assumed (bottom right). The most probable value of the energy distribution (MPV), indicated as circles (top left), is shifted towards lower beam energies with increasing $\sigma_\mathrm{TOF}$. The symmetric 1-$\sigma$ interval of the TOF distributions are highlighted as dotted lines and the corresponding energy values as dashed lines. }
\label{fig:tofdistort}
\end{center}
\end{figure}

\begin{figure}[!h]
  \begin{center}
  \includegraphics[width=0.99\textwidth]{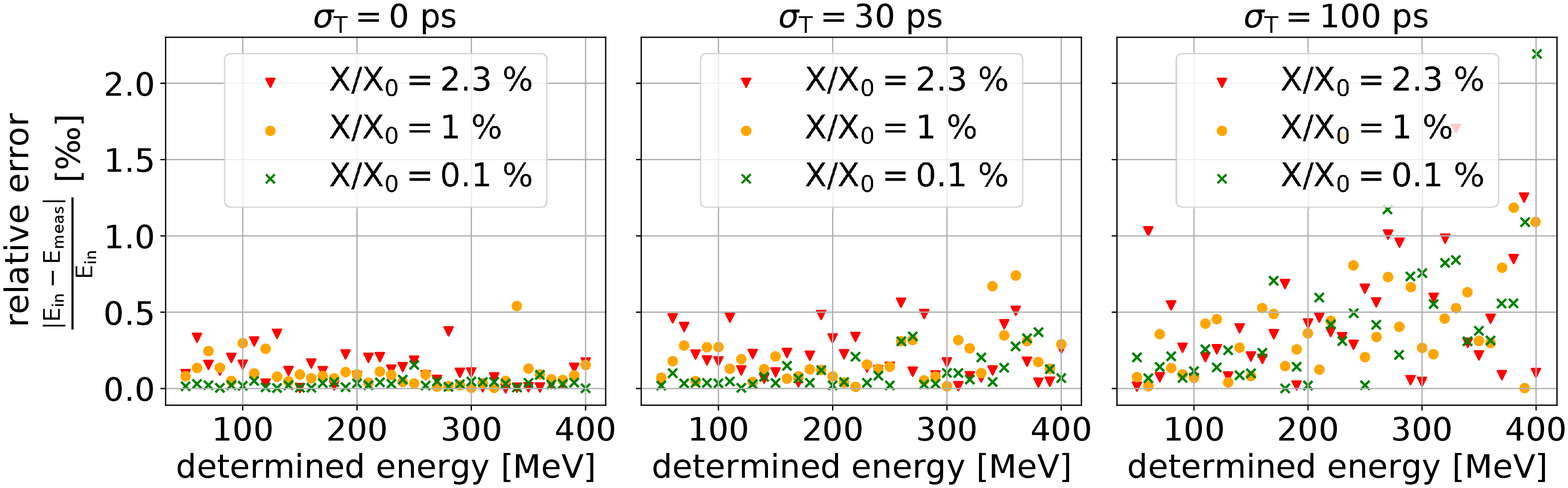}
\caption{Accuracy of the energy measurement for an LGAD based TOF calorimeter after calibration. The relative error is shown for a setup with a flight distance of $\SI{0.5}{m}$, different material budgets, intrinsic time resolutions per tracker plane and determined residual proton energies ($E_{\mathrm{meas}}$). }
\label{fig:postcalib}
\end{center}
\end{figure}

\mbox{}\\
\label{sec:resultcalib}
The importance of an energy calibration for the TOF calorimeter can be seen when looking at figure \ref{fig:praecalib}, where the absolute error of the energy measurement is shown for different system settings. 
For an ideal setup with $\SI{0}{ps}$ time resolution, the absolute error increases with increasing material budget per LGAD module and decreasing beam energy due to the significant energy loss inside the detectors. At higher beam energies, the energy loss decreases due to lower stopping power values, and therefore smaller absolute errors are expected. However, when realistic intrinsic time resolutions are assumed ($\geq \SI{30}{ps}$), an increase of the absolute error at higher beam energies can be observed. The reason for this increased inaccuracy at higher beam energies is shown in figure \ref{fig:tofdistort}, where the theoretical time-of-flight is depicted for $\SI{800}{MeV}$ protons and a flight distance of $\SI{1}{m}$ (lower left). Since this effect is less pronounced at lower beam energies, $\SI{800}{MeV}$ protons were used to better illustrate this behaviour. As depicted in figure \ref{fig:tofdistort}, due to the non-linear relation between TOF and kinetic energy (equation (\ref{eq:evstof})), the most probable value (MPV) of the measured kinetic energy distribution (top left) is shifted towards lower beam energies if uncertainties of the TOF measurement are assumed (bottom right). In addition, the energy distribution is skewed, with a large tail towards higher beam energies. To show this effect, the theoretical TOF was distorted using a Gaussian distribution with a standard deviation of $\sigma_{\mathrm{TOF}}$ emulating an uncertainty of the TOF measurement while neglecting the energy loss along the flight path. Using $\SI{30}{}$ and $\SI{100}{ps}$  resulted in a shifted MPV (circles top left) of $\SI{797}{MeV}$ and $\SI{768}{MeV}$, respectively. In order to highlight the asymmetry of the resulting energy distribution, the corresponding energies of the symmetric 1-$\sigma$ interval of the TOF distribution (dotted lines) are shown as dashed lines. 
	 \mbox{}\\
To correct for the inaccuracies in the energy measurement, a calibration as described in equation (\ref{eq:calibcurve}) has to be applied. Therefore, an empirical model $f(E)$ 
\begin{equation}
 = \frac{{(E-a)}^{1-b}}{d\cdot b}-\frac{e\cdot  m_0 c^3 {\gamma(E)}^3 {\beta(E)}^2}{f} \left(\frac{g}{c}-\beta(E)\cdot h \right)
\label{eq:calibrationcurve}
\end{equation}
was fitted to the observed absolute errors of the energy measurement for each setting. The first part of equation (\ref{eq:calibrationcurve}) is based on the energy loss of a particle in a homogeneous medium. In general, the total energy loss inside a realistic TOF calorimeter is more complicated since it is the result of multiple energy losses at different parts of the calorimeter, which strongly depends on the setup geometry. The second part is a 2D Taylor expansion of equation (\ref{eq:evstof}) to account for inaccuracies in the TOF and path length estimation. However, it has to be mentioned that the calibration curve in equation (\ref{eq:calibrationcurve}) is not an exact analytical model of the absolute error of the energy measurement, but provides a robust parametrization of the calibration curves for all investigated detector geometries.\\
After applying the resulting calibration curves, highlighted as dashed lines in figure \ref{fig:praecalib}, to a new set of simulations, the relative error $\epsilon_E$ (equation (\ref{eq:relerrenergy})) was calculated for each TOF calorimeter configuration to estimate the accuracy of the energy measurement. For all investigated settings, only relative errors below $\SI{0.22}{\percent}$ could be observed. As an example, the obtained accuracy values for a TOF calorimeter with a flight distance of $\SI{0.5}{m }$  and different system parameters are shown in figure \ref{fig:postcalib}.

\subsubsection{Influence of the spatial resolution}\mbox{}\\
The influence of the intrinsic spatial resolution on the energy resolution and accuracy of a stand-alone TOF calorimeter was studied for $\sigma_{xy}$ ranging from $\SI{0}{cm}$ to $\SI{5}{cm}$. Below $\sigma_{xy}=\SI{1}{cm}$, no impact of the spatial resolution on the performance of the energy measurement could be observed for all investigated system settings. For settings with $\sigma_{xy}<\SI{1}{cm}$, the accuracy and the energy resolution only differed by $\leq \SI{0.05}{\percent}$ and $\leq \SI{0.033}{\percent}$, respectively, when compared to  $\sigma_{xy}=\SI{0}{cm}$. Consequently, the intrinsic spatial resolution should only affect the tracking performance of the proposed TOF-pCT system, which uses the same detector technology for particle tracking and residual energy measurements. The influence of the intrinsic spatial resolution of tracking detectors on the MLP estimation and achievable image voxel sizes has already been discussed extensively in \cite{Krah_2018,burker2021single}. Therefore, since this work focuses on the TOF-based residual energy measurement, the spatial resolution was set to zero for all of the following simulations of different TOF-pCT systems.

\subsection{Proton computed tomography system} 
\begin{figure}[!h]
 	\begin{center}
 	\includegraphics[width=0.9\textwidth]{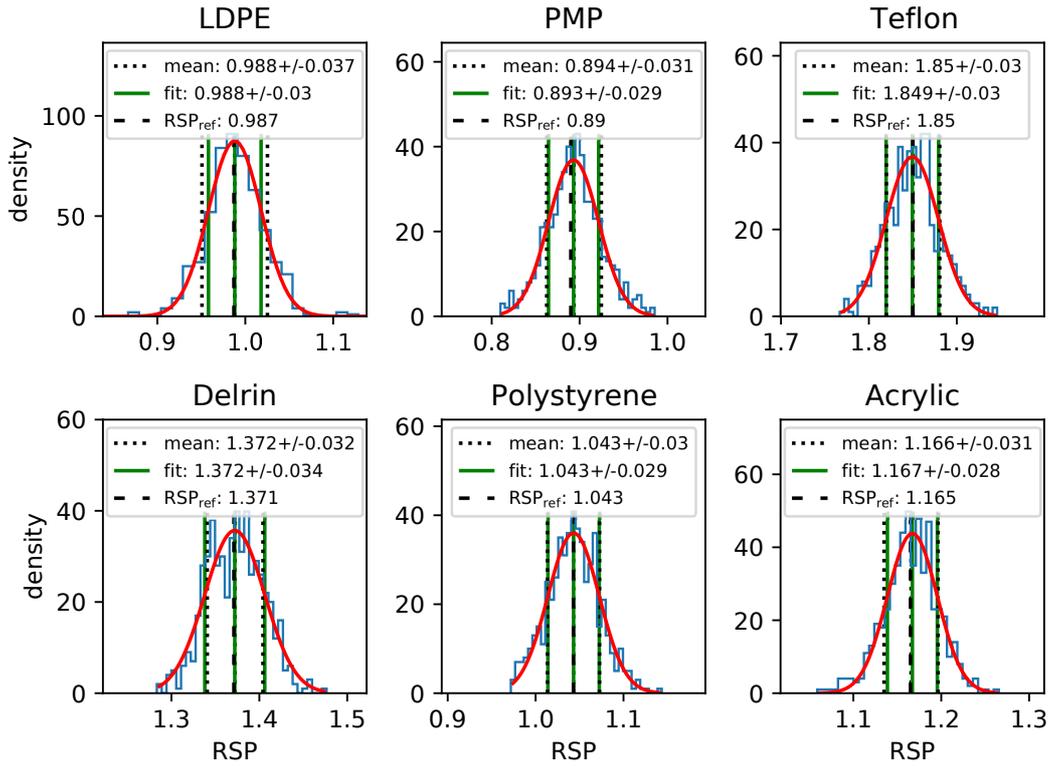}
 	\end{center}
 	\caption{Collected RSP distributions in the CTP404 inserts (blue histograms) for $\SI{200}{MeV}$ protons, a flight distance of $\SI{1}{m}$, $\SI{0.1}{\percent} \mathrm{X}/\mathrm{X}_0$ and $\SI{30}{ps}$ $\sigma_{\mathrm{T}}$ per LGAD module. The sample RSP mean and sample RSP standard deviation (dotted lines), as well as the RSP mean and standard deviation obtained from a Gaussian fit (solid lines), are compared to reference RSP values (dashed lines). For almost all of the depicted inserts, the sample mean, fitted mean and reference value overlap. }
 	\label{fig:insertshisto}
 	\end{figure}

As discussed in section \ref{sec:phantom}, the RSP was measured inside the inserts of the CTP404 phantom to estimate the performance of the TOF-based pCT system. An example for the acquired RSP distributions per insert is shown in figure \ref{fig:insertshisto}. Also, the theoretical RSP values, obtained via a simulated residual range measurement, are listed. The sample mean of the RSP, as well as the sample standard deviation are in good agreement with the RSP values obtained from the Gaussian fit and the reference RSP values. 
\subsubsection{RSP precision}
\begin{figure}[!h]
  \begin{center}
  \includegraphics[width=0.9\textwidth]{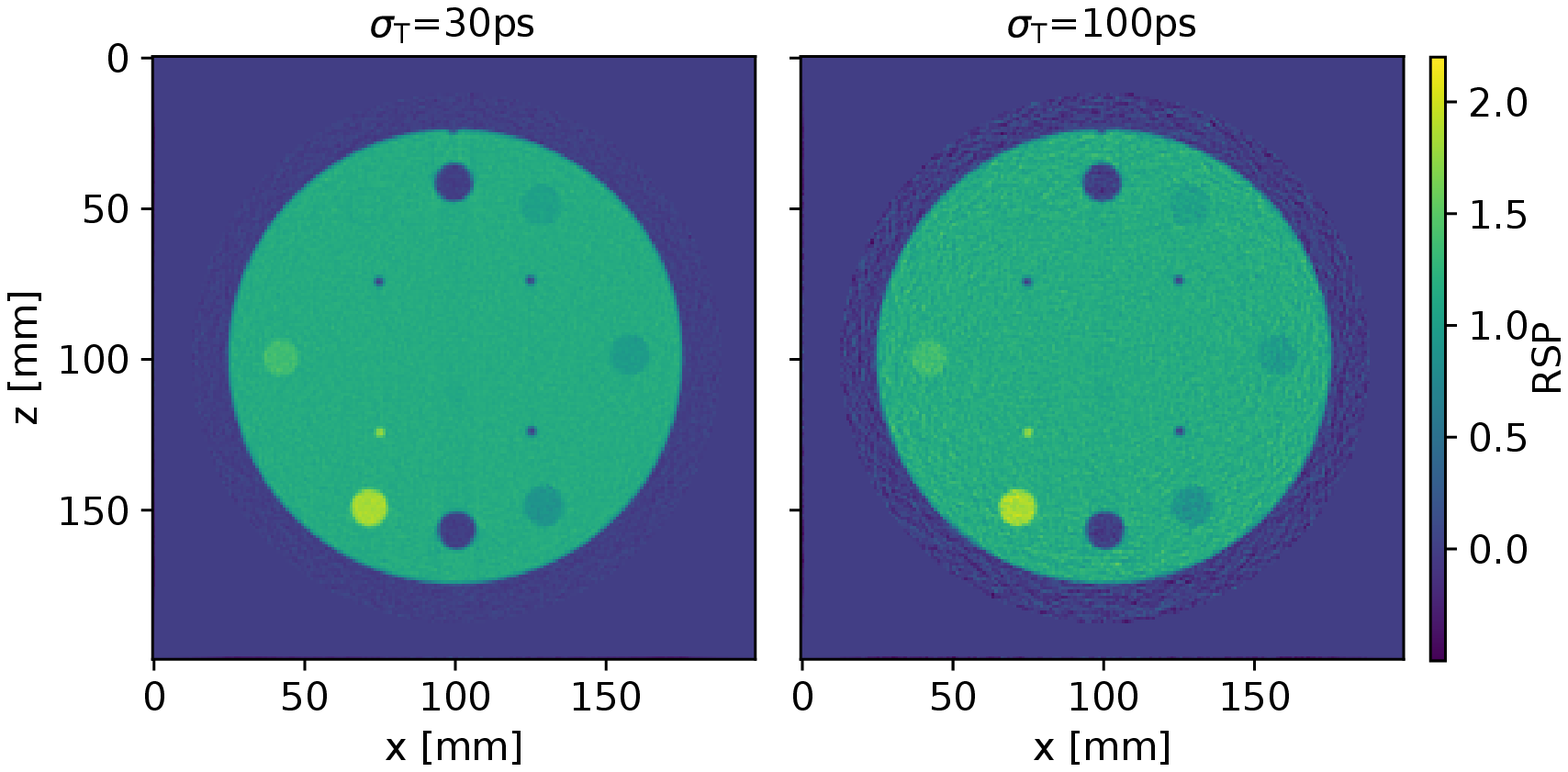}

\caption{Reconstructed central slices of the CTP404 phantom recorded with $\SI{200}{MeV}$ protons and $\sigma_\mathrm{T}=\SI{30}{ps}$ (left) and  $\sigma_\mathrm{T}=\SI{100}{ps}$ (right). The pixels outside the field of view (FOV) have been set to 0.}
\label{fig:allslices}
\end{center}
\end{figure}

\begin{figure}[!h]
  \begin{center}
\includegraphics[width=0.99\textwidth]{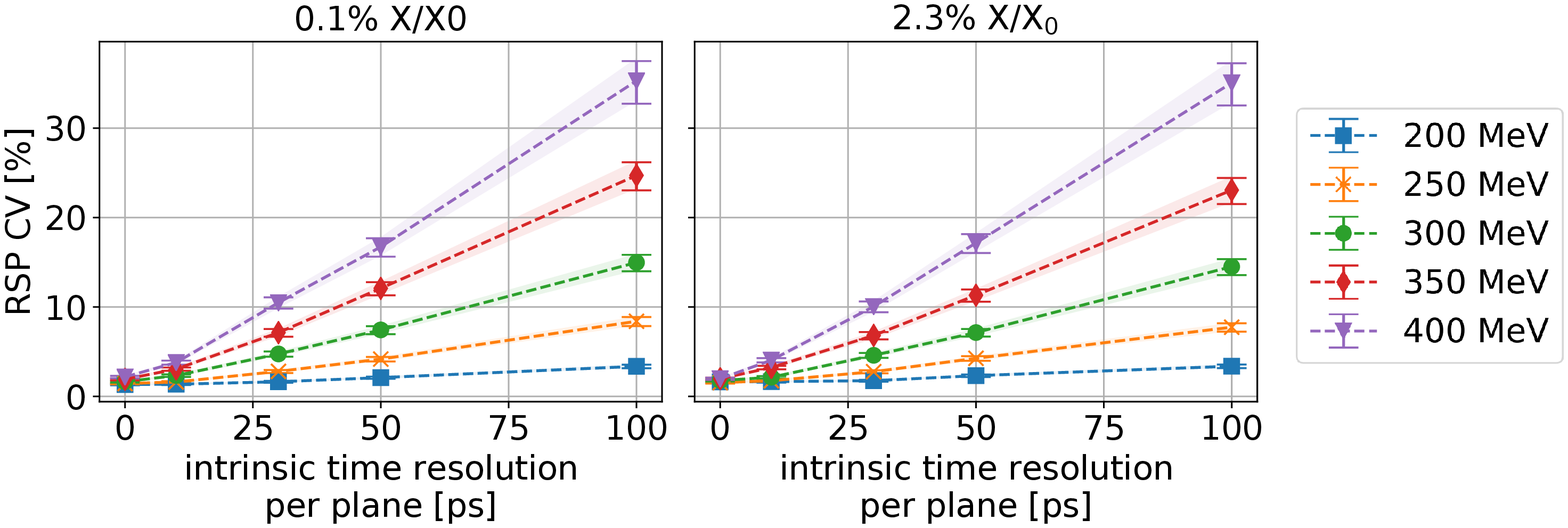}
\label{fig:teflonCV}
\caption{Energy dependence of the RSP precision. The RSP CV, measured in the Teflon insert, is shown for different beam energies and intrinsic time resolutions $\sigma_{\mathrm{T}}$. The results were obtained using a TOF-pCT system with $\SI{0.1}{\percent} \mathrm{X}/\mathrm{X}_0$ (left) and $\SI{2.3}{\percent} \mathrm{X}/\mathrm{X}_0$ (right) LGAD planes and a flight distance of $\SI{1}{m}$. The error bars and error bands represent the $\SI{95}{\percent}$ confidence interval for the CV, which was calculated according to \cite{McKay1932}.}
\label{fig:CVsingle}
\end{center}
\end{figure}

\begin{figure}[!h]
  \begin{center}
\includegraphics[width=0.99\textwidth]{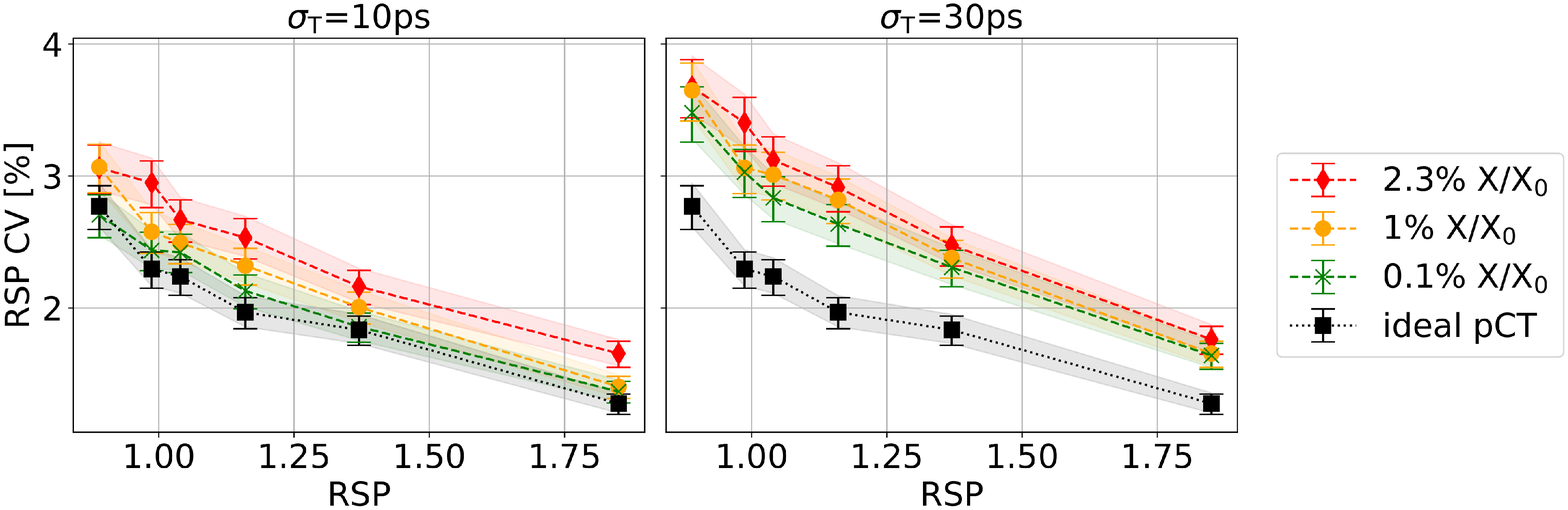}
\caption{Impact of the material budget on the RSP precision. The RSP coefficient of variation was measured in all six inserts ($x$-axis) using 200 MeV protons. All depicted TOF-CT systems were simulated with a flight distance of 1m and were compared to an ideal pCT simulation. The error bars and error bands represent the $\SI{95}{\percent}$ confidence interval for the CV, which was calculated according to \cite{McKay1932}. }
\label{fig:rspprec}
\end{center}
\end{figure}

\mbox{}\\
Similar to the energy resolution of the TOF calorimeter, the RSP precision strongly depends on the intrinsic time resolution per tracking plane and the beam energy. This also becomes evident when looking at figure \ref{fig:allslices}, where the central slices of the reconstructed CTP404 phantom are shown for $\SI{30}{}$ and $\SI{100}{ps}$ intrinsic time resolutions per tracking plane. The noise for $\sigma_\mathrm{T}=\SI{100}{ps}$ (right) is increased in contrast to $\sigma_\mathrm{T}=\SI{30}{ps}$ (left) due to the inferior intrinsic time resolution. With increasing beam energy and inferior intrinsic time resolutions, even more noise was observed in the central slices.
 \mbox{}\\
In order to quantify the RSP precision, the CV of the RSP was measured in each insert. As an example, the obtained CV for Teflon is shown in figure \ref{fig:CVsingle} for a pCT system with a flight distance of $\SI{1}{m}$, $\SI{0.1}{\percent}$ (left) and $\SI{2.3}{\percent}$ (right) $\mathrm{X}/\mathrm{X}_0$. In general, the RSP precision improves with decreasing beam energy and increased intrinsic time resolution. To better visualize the dependence of the RSP precision on the material budget, the CV values, measured in all six inserts, are depicted in figure \ref{fig:rspprec} for all investigated material budgets, $\SI{200}{MeV}$ protons and a flight distance of $\SI{1}{m}$. The CV increases only slightly with increasing material budget due to additional straggling in the detector planes. However, as observed in figure \ref{fig:CVsingle}, the intrinsic time resolution affects the RSP precision more dominantly. For example, if an intrinsic time resolution of $\SI{10}{ps}$ is assumed (figure \ref{fig:rspprec}), the CVs are much closer to the values obtained from the ideal pCT simulation, where the RSP precision is only dominated by the straggling inside the phantom. Especially for the LGADs with the lowest investigated material budget, the difference of the measured CV values could be decreased to $\approx \SI{2}{\percent}-\SI{9}{\percent}$.  

\subsubsection{RSP accuracy}
\begin{figure}
  \begin{center}
\includegraphics[width=0.99\textwidth]{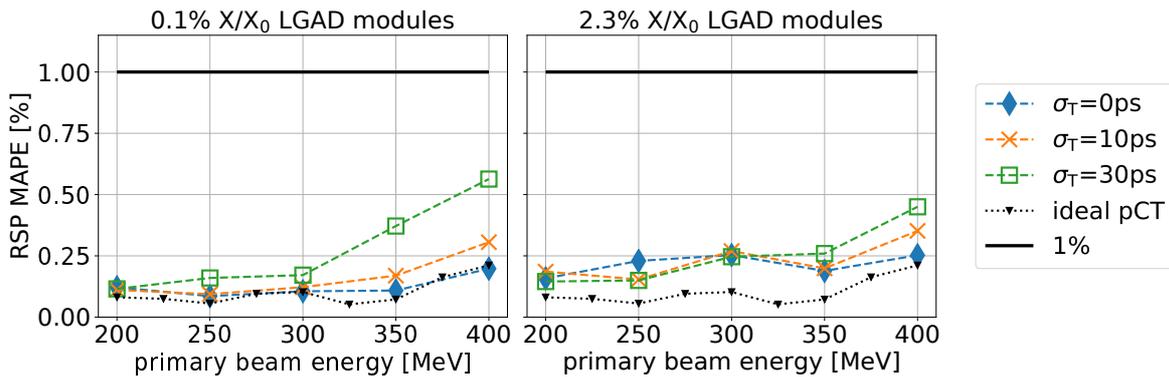}
\caption{RSP MAPE for different material budgets and different intrisic time resolutions per tracking plane at $\mathrm{D}_{\mathrm{TOF}}=\SI{1}{m}$. The MAPE was calculated for different primary beam energies and compared to a simulation of an ideal pCT setup. After calibrating the TOF-pCT systems, the MAPE was always well below the $\SI{1}{\percent}$ requirement \cite{schulte2004} for all investigated settings.}
\label{fig:mapes}
\end{center}
\end{figure}

\mbox{}\\
Figure \ref{fig:mapes} illustrates the measured RSP MAPEs for different pCT system parameters. For all investigated settings, the proposed calibration procedure allowed to achieve RSP accuracies close to the theoretical limit, defined by the simulation of the ideal pCT system. The obtained RSP MAPEs varied between $\SI{0.12}{}$ and $\SI{0.6}{\percent}$, which is well below the requirements for a clinical pCT system \cite{schulte2004}. Also, the lower limit of the obtained MAPE is in good agreement with the MAPE obtained from an ideal pCT system simulated in \cite{Dedes2019} and only differed by $\approx \SI{0.1}{\percent}$, which could result from a different estimation of the reference RSP values, as described in section \ref{sec:rsp}. As depicted in figure \ref{fig:mapes}, the ideal setup and the investigated TOF-pCT systems showed similar fluctuations in the obtained MAPE, which depend on the beam energy. Only for $\SI{0.1}{\percent}$ $\mathrm{X}/\mathrm{X}_0$, $\sigma_\mathrm{T}$=$\SI{30}{ps}$ and beam energies $\geq\SI{350}{MeV}$, considerable differences were observed, which could indicate a non-ideal energy calibration for those specific settings. \\
For a more detailed comparison of the investigated TOF-pCT setups, the relative RSP errors per insert, the standard errors of the mean, as well as the MAPEs are listed in table \ref{tab:mapes} for TOF-pCT systems with $\SI{30}{ps}$ intrinsic time resolution, different material budgets per LGAD module and a primary beam energy of $\SI{200}{MeV}$.
 	
\begin{table}[!htp]
\centering
\setlength\tabcolsep{0pt}

\begin{tabular*}{\textwidth}{
  @{\extracolsep{\fill}}
  l 
  S[table-format=-2.3(5)] |
  S[table-format=-2.3(6)]
  S[table-format=-2.3(6)]
  S[table-format=-2.3(6)]
  @{}
}
\toprule 
&\hspace{0.8cm}\textbf{ideal pCT}& \multicolumn{3}{c}{\textbf{TOF-pCT with \boldmath{${\sigma_\mathrm{T}}={{30}{\hspace{0.1cm}\mathrm{ps}}}$}}}  \\
\boldmath{$\frac{\mathrm{X}}{\mathrm{X}_0}$} ${\left[ \% \right]}$ & \hspace{1.6cm}0& \hspace{1.6cm}0.1&\hspace{1.6cm}1&\hspace{1.6cm}2.3\\
\midrule
\textbf{PMP} &0.232\pm 0.119 & 0.410\pm 0.150 & 0.306\pm 0.158 &-0.033\pm 0.16  \\
\textbf{LDPE} &-0.004\pm 0.099 & 0.098\pm 0.162 & 0.177\pm 0.132 &0.262\pm 0.153  \\
\textbf{Polystyrene} & -0.030\pm 0.096 & 0.012\pm 0.122 & 0.007\pm 0.120 &0.211\pm 0.135  \\
\textbf{Acrylic} & 0.035\pm 0.085 & 0.057\pm 0.113 & 0.162\pm 0.121 &0.154\pm 0.133  \\
\textbf{Delrin} & -0.33\pm 0.079 & 0.103\pm 0.099 & 0.074\pm 0.102 &-0.008\pm 0.107  \\
\textbf{Teflon} & -0.153\pm 0.055 & 0.011\pm 0.071 & -0.007\pm 0.712 &-0.202\pm 0.098  \\
\midrule
\textbf{MAPE} \boldmath{${\left[ \% \right]}$} & \hspace{1.2cm}\textbf{{0.081}} &\hspace{1cm} \textbf{{0.115}} &\hspace{1cm} \textbf{{0.122}} &\hspace{1cm} \textbf{{0.145}}\\
\bottomrule
\end{tabular*}
  \caption{Relative RSP errors ${\left[ \% \right]}$ of the pCT system for 200 MeV protons, a flight distance of 1m and an intrinsic time resolution of $\SI{30}{ps}$. The standard error of the mean was used to estimate the uncertainty of the RSP accuracy in each insert.}
  \label{tab:mapes}
\end{table}

\section{Discussion}
The aim of this work was to investigate the feasibility of a pCT system based on 4D-tracking detectors with a TOF calorimeter for the residual energy measurement. Therefore a realistic pCT system was modelled in Geant4, and various design parameters were varied and optimized. In general, the proposed pCT system could also be used for other ion species. The advantages of using a TOF system for He-CT have already been discussed in \cite{volzphd2021, Rovituso2017}. For example, in \cite{volzphd2021}, the performance of a theoretical TOF-based residual energy calorimeter for helium ion computed tomography (HeCT) has been studied in terms of RSP resolution. The authors of \cite{Rovituso2017}, on the other hand, have shown that measuring the TOF through the patient could be used for particle identification to remove nuclear interaction events. However, since investigating other ions species would go beyond the scope of this study, we have only presented the results for a TOF-based proton computed tomography system.
\subsection{Impact of the investigated system parameters on the energy measurement}
The performance of a stand-alone TOF-based residual energy calorimeter has been investigated in terms of precision and accuracy. \\To achieve WET resolutions close to the theoretical range straggling limits, the energy resolution of a residual energy calorimeter (single staged) for pCT should be $<\SI{1}{\percent}$ for residual energies ranging from a few tens of $\SI{}{MeV}$ to a few hundreds of $\SI{}{MeV}$ \cite{Bashkirov2016}. Thus, it was necessary to identify and discuss the design parameters that influence the energy resolution of the TOF calorimeter prior to designing a realistic pCT system based on 4D-tracking detectors. Similar results as described in \cite{volzphd2021} were found, where the influence of the intrinsic time resolution and flight distance of a TOF calorimeter on the precision of an iCT with $\SI{200}{MeV/u}$ He-ions was studied. The study presented in this work, on the other hand, also investigated the influence of spatial resolution, material budget, and different residual beam energies for a more comprehensive analysis, which should serve as a guide for future hardware developments. As indicated in figure \ref{fig:eresresults}, the energy resolution improves with decreasing residual beam energy down to $\approx \SI{100}{MeV}$ for all investigated settings. However, at lower energies, depending on the setting, a significant contribution due to increased energy straggling inside the TOF calorimeter could be observed. Therefore, to fully optimize the residual beam energy for each setting, a simulation of the calorimeter setup should be preferred over the analytical model (equation (\ref{eq:eres})), where energy straggling was not taken into account.
The energy resolution is also strongly impacted by the intrinsic time resolution per plane. In general, LGADs with the most precise intrinsic time resolution should be used for the TOF calorimeter to improve the energy resolution. Recent studies have shown that time resolutions of $\approx\SI{30}{ps}$ can be achieved \cite{CARTIGLIA201783}. For an LGAD based TOF calorimeter with $\sigma_{\mathrm{T}}=\SI{30}{ps}$ and a flight distance of $\SI{1}{m}$, residual energies should be $\leq \SI{100}{MeV}$ in order to reach the desired energy resolutions below the $\SI{1}{\percent}$ limit. Alternatively, longer flight distances could be used for the same intrinsic time resolution when higher residual beam energies are expected.\\
The energy loss inside the detector modules and the intrinsic time resolution have been identified as the main sources of inaccuracy for the energy measurement. In order to account for those inaccuracies, a dedicated calibration procedure has been introduced. After applying the calibration, the relative error of the energy measurement could be decreased to $\approx\SI{0.2}{\percent}$ for all settings. Within the investigated parameter space, no significant influence of the flight distance, beam energy, material budget or intrinsic time resolution on the accuracy of the energy measurement could be observed after the calibration.
\\ 
Another important design aspect is the granularity of the 4D-tracking detector. The concept of high-granularity detectors for pCT has already been introduced in \cite{Alme2020} to cope with the high particle rates of clinical pencil beams. Using detectors with high granularity reduces the sensor occupancy since fewer particles will traverse the same detector cell at the same time. This allows recording a high multiplicity of incoming particle tracks, which results in an increased rate capability. Within this work, we investigated whether additional limitations on the granularity have to be imposed in terms of required spatial resolution for the residual energy measurement. However, for $\sigma_{xy}<\SI{1}{cm}$, no significant influence on the performance of the energy measurement could be observed. Consequently, the design choice for the sensor granularity of the LGAD based TOF-pCT system is mainly driven by the required image voxel size \cite{Krah_2018, burker2021single} and rate capability of the 4D-tracking system.
\subsection{Impact of the investigated system parameters on the RSP determination}
As described in \cite{schulte2004}, a clinical pCT system should be able to measure the RSP inside $\leq\SI{1}{mm^3}$ voxels with an accuracy better than $\SI{1}{\percent}$. Those requirements have already been fulfilled by the phase II preclinical pCT prototype scanner \cite{JOHNSON2017209}, which achieved RSP accuracies of $\approx \SI{0.69}{\percent}$ \cite{Dedes2019}. To show the potential of pCT to further improve the RSP accuracy, the authors of \cite{Dedes2019} have also simulated an ideal pCT system with infinitesimally thin detectors and ideal energy and position measurements, which reached RSP accuracies down to $\approx \SI{0.17}{\percent}$. Following the example of \cite{Dedes2019}, we also simulated an ideal pCT system for verification and for comparing the performance of the realistic TOF-based pCT scanner. Similar to the stand-alone TOF calorimeter, no significant dependence on any system parameter on the RSP accuracy was observed after applying the calibration as described in the previous sections. Only for higher beam energies, using higher material budgets, the MAPE increased up to $\SI{0.6}{\percent}$, which is still well below the $\SI{1}{\percent}$ margin. The best RSP accuracy ($\SI{0.12}{\percent}$) was obtained for a pCT system with $\SI{0.1}{\percent} \mathrm{X}/\mathrm{X}_0$ LGADs, a flight distance of $\SI{1}{m}$ and intrinsic time resolution of $\SI{30}{ps}$, which is close to the theoretical limit defined by the ideal pCT simulation (table \ref{tab:mapes}). The promising improvements in terms of RSP accuracy indicate that a TOF-based pCT scanner could outperform the latest DECT scanners used for treatment planning \cite{Dedes2019}, which has to be confirmed with an experimental prototype.
\\
Figure \ref{fig:eresresults} suggests that if $\SI{10}{ps}$ intrinsic time resolution per LGAD is assumed, the resulting WET resolution should be close to the theoretical straggling limit, independent of the investigated flight distances and residual beam energies. Similarly, the RSP precision was also always close to the results obtained from the ideal pCT simulation if $\sigma_{\mathrm{T}}$ was set to $\SI{10}{ps}$. Assuming a more realistic intrinsic time resolution (e.g. $\SI{30}{ps}$) resulted in a significant decrease of RSP precision as indicated in figure \ref{fig:CVsingle} and \ref{fig:rspprec}. However, the system parameters were not fully optimized to boost the RSP precision. In general, for a realistic LGAD system with $\sigma_\mathrm{T}\geq\SI{30}{ps}$, the residual beam energy should be kept as low as possible as indicated in figure \ref{fig:eresresults}. Also, depending on the available space in the treatment room, the flight distance can be adapted according to equation (\ref{eq:planes}) to improve the energy resolution and, therefore, the RSP precision. Alternatively, as described in equation (\ref{eq:planes}), using more LGADs per timing station could improve the energy resolution. However, since this would increase the cost of the pCT system, optimizing the other system parameters should be preferred.

    \section{Conclusion}
    The main purpose of this study was to offer a comprehensive overview of the most important system parameters of a realistic TOF-pCT scanner based on 4D-tracking detectors, serving as a guide for future hardware developments. Using MC simulations, we could demonstrate that a TOF-based pCT system could potentially achieve RSP accuracies well below the $\SI{1}{\percent}$ margin, if properly calibrated, and therefore improve the treatment plan quality. Using a model of a realistic TOF-pCT setup with $\SI{1}{\percent}$ $\mathrm{X}/\mathrm{X}_0$, $\SI{30}{ps}$ intrinsic time resolution and a flight distance of $\SI{1}{m}$ resulted in an RSP accuracy of $\approx \SI{0.12}{\percent}$, which could outperform the latest DECT scanners ($\approx \SI{0.6}{\percent}$). To achieve an RSP resolution close to the straggling limit, intrinsic time resolutions of at least $\SI{30}{ps}$ are recommended. However, for a more realistic setup with time resolutions $\geq\SI{30}{ps}$, system parameters such as residual beam energy, flight distance and number of LGADs should be optimized to further improve the RSP precision.
    
    \section*{Acknowledgements} 
    This project received funding from the Austrian Research Promotion Agency (FFG), grant number 869878.

    \section*{References}
    \bibliography{tofcalct}

\end{document}